\documentclass[fleqn,12pt,twoside]{article}
\usepackage{espcrc1}

\usepackage{graphicx}

\def\pmb#1{\setbox0=\hbox{#1}%
     \kern-.025em\copy0\kern-\wd0
      \kern.05em\copy0\kern-\wd0
       \kern-.025em\raise.0433em\box0}

\date{\today}
\title
{Dynamics of Phase Transitions in Asymmetric Nuclear Matter}
\author{
M.Di Toro\address[LNS]{Laboratori Nazionali del Sud INFN, Via S. Sofia 44,
I-95123 Catania, Italy\\
and Dipartimento di Fisica, Universit\`a di Catania},
 V. Baran\addressmark[LNS]%
\thanks{On leave from NIPNE-HH and Bucharest University, Romania},
 M.Colonna\addressmark[LNS],
 A.Drago\address{Dip.di Fisica, Univ. di Ferrara and INFN Sezione di Ferrara},
 T. Gaitanos\addressmark[LNS]%
\thanks {On leave from Sektion Physik, 
 Universit\"at M\"unchen, Germany, supported by the BMBBW grant 06LM981}
V.Greco\address{Cyclotron Institute, Texas A\&M Univ., College Station, USA}%
\thanks{On leave from Laboratori Nazionali del Sud INFN, Catania, Italy
and Dipartimento di Fisica, Universit\`a degli Studi di Catania, partially
supported by a grant of the A.LaRiccia Foundation},
and
A.Lavagno\address{Dip.di Fisica, Politecnico di Torino and INFN 
 Sezione di Torino}
}

\begin{document}

\maketitle

\begin{abstract}
We present several possibilities offered by the reaction dynamics 
of dissipative heavy ion collisions to study in detail the symmetry
term of the nuclear equation of state, $EOS$. In particular
we discuss isospin effects on the nuclear liquid-gas phase transition, 
{\it Isospin Distillation}, and on collective flows.
We stress the importance of a microscopic relativistic structure
of the effective interaction in the isovector channel.
The possibility of an {\it early} transition to deconfined matter
in high isospin density regions is also suggested.
We finally select {\it Eleven} observables, in different beam energy regions,
 that appear rather sensitive to 
the isovector part of the
nuclear $EOS$, in particular in more exclusive experiments.

\end{abstract}

\vskip 1.0cm

\subsection*{Introduction}

There are quite stimulating predictions on new 
phases of Asymmetric Nuclear Matter, $ANM$,  
that eventually could be reached during
heavy ion reaction dynamics with radioactive beams
 \cite{iso,mue95,bao197,cdl98,bar98,bar01} .
More symmetric
and narrower isotopic
distribution of primary fragments are predicted, {\it and 
sensitive to the symmetry term of the $EOS$}.
For semi-central collisions the dynamics of the participant
zone appears also to be quite affected by the symmetry term 
\cite{fab98,eri98,neck,ditcr,bar02}.

Collective flows are particularly interesting since we can
probe different density regions of the $EOS$. A very stimulating result 
shown here
is the sensitivity to the microscopic covariant structure
of the isovector channel in the {\it in medium} interaction.
 A related {\it earlier} possible
transition to a mixed phase with deconfined matter is finally
presented.

\noindent
{\it The EOS symmetry term}

The behaviour of the symmetry term of the nuclear $EOS$ is poorly known  
in regions far from normal density. 
In the following we will compare results obtained with
forces that have  {\it the same saturation properties for
symmetric $NM$}.
We will refer to a "asy-stiff/superstiff" $EOS$ when we are 
considering a potential symmetry term with a linear/parabolic 
increase with nuclear 
density and to a "asy-soft" $EOS$ when it shows a 
saturation and eventually a decrease above normal density \cite{eri98}.

\begin{figure}
\begin{minipage}{75mm}
\begin{center}           
\includegraphics*[scale=0.36]{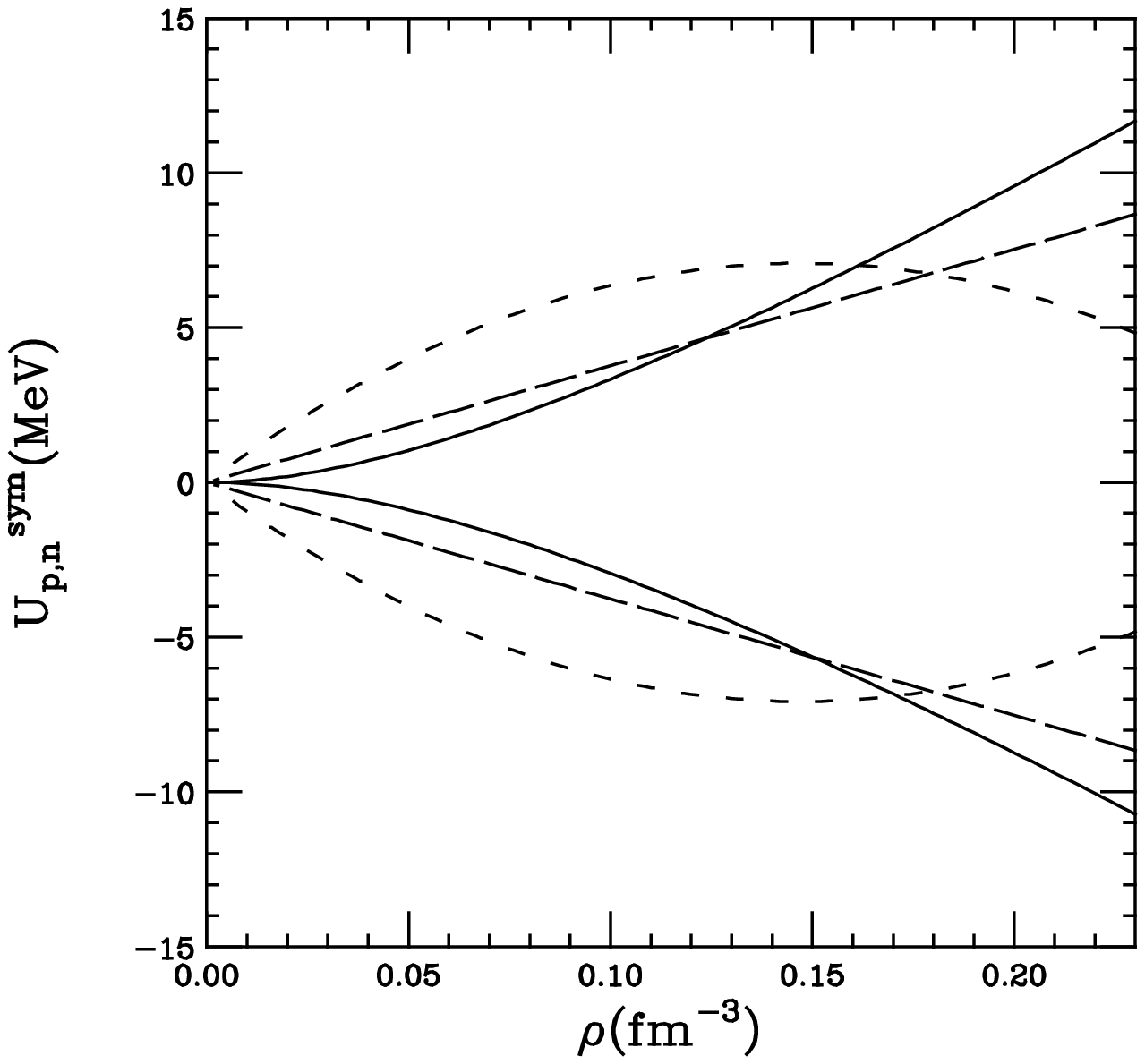}
\caption
{\it Symmetry contribution to the mean field at $I=0.2$
for neutrons and protons:
dashed lines "asy-soft", solid
lines "asy-stiff", long dashed lines "asy-superstiff"}
\label{mean}
\end{center}
\end{minipage}
\hspace{\fill}
\begin{minipage}{75mm}
\begin{center}
\includegraphics*[scale=0.35]{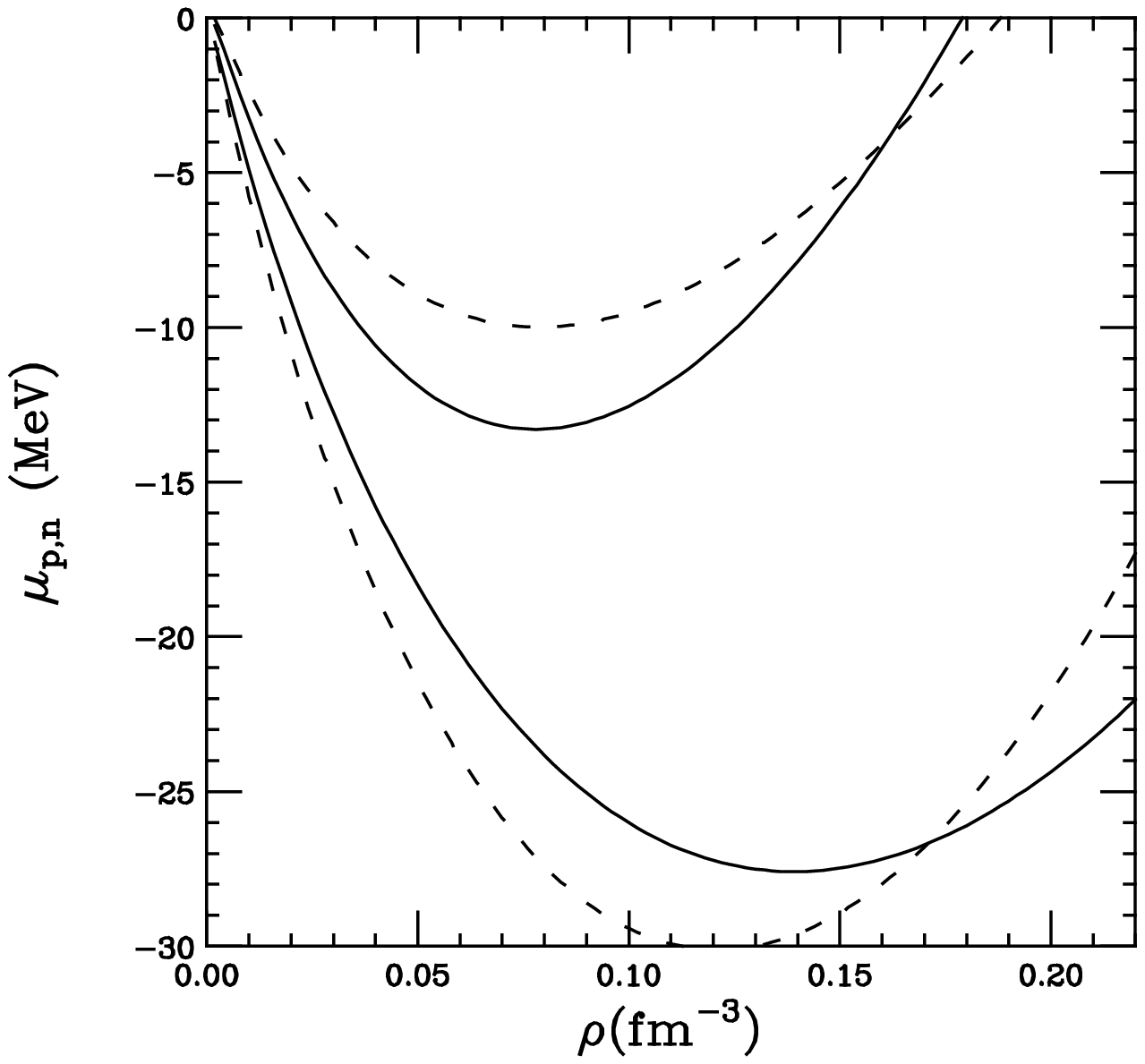}
\caption{\it Density dependence of proton and neutron chemical potential
for asy-superstiff (solid lines) and asy-soft (dashed line) EOS.}
\label{chem}
\end{center}
\end{minipage}
\end{figure}

In Figs.\ref{mean},\ref{chem} we report, for 
a $^{124}Sn$ asymmetry $(N-Z)/A=0.2$, 
the density dependence of the 
symmetry contribution to the mean-field potential (left) and of the
chemical potentials (right) for neutrons (top curves)
and protons (bottom curves) , for the different 
effective interactions in the isovector channel.  
From the behavior in the low densities region we expect that
when the inhomogenities develop both neutrons
and protons have the tendency to move from lower to higher density
regions, in phase: the system is unstable
against {\it isoscalar-like} fluctuations and not isovector, see later.
Since
the variations of the two chemical
potentials are different (larger for protons) we expect a lower asymmetry in
the liquid phase. 
In the case of a contact between 
more dilute and  "normal" density regions, we see from Fig.\ref{chem}
that in this range the neutrons have 
the tendency to move towards the dilute part producing a $n$-enrichment
while the protons will migrate to the higher density 
regions. This mechanism is present in the "neck fragmentation",
\cite{neck,ditcr,bar02}:
the neck $IMF's$ will be always more $n-$rich compared to the fragments
produced in
the case of bulk fragmentation. 

Stochastic transport simulations \cite{fab98,gre98,flows,flu98}
of fragment production collisions
at medium energies are confirming these predictions, see refs. 
\cite{ditcr,bar02} where a comparison with recent data \cite{ts01,chimera}
is also performed. 

\subsection*{Isospin Distillation in Dilute Nuclear Matter}
For charge asymmetric systems we expect a qualitative new feature
in the liquid-gas phase transition, the onset of chemical instabilities
that will show up in a novel structure of the unstable modes
\cite{bar98,bar01}.
Experimentally this will be revealed through the $Isospin~Fractionation$
or $Distillation$ effect in multifragmentation events.
\begin{figure}
\begin{center}
\includegraphics*[scale=0.35]{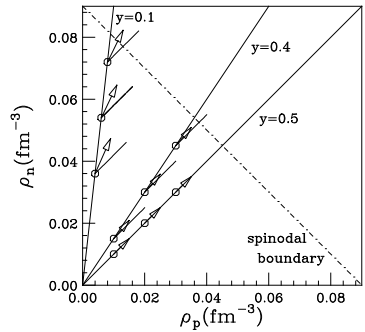}
\caption{\it The mechanism of isospin distillation. The arrows show
the ``direction'' of the unstable mode in various points of the
spinodal region. The thick lines represent a composition corresponding
to the initial concentration, the thin lines a pure isoscalar mode.
Initial proton fractions $y=0.5,~y=0.4~and~y=0.1$}
\label{modes}
\end{center}
\end{figure}
We have now a new degree of freedom, the concentration, and 
in the spinodal region 
the fluctuations against which a binary
system becomes unstable depend on the nature of the
interaction between the two components of the mixture.
We define density fluctuations as isoscalar-like in 
the case when proton and neutron densities fluctuate in phase and as
isovector-like  when the oscillations are
out of phase. 
For the dilute asymmetric nuclear
matter because of the attractive force between protons and neutrons 
at low density
the phase transition is {\it uniquely} driven by isoscalar-like 
instabilities \cite{bar01,cho03}. 

An intuitive picture is presented in Fig.\ref{modes}.
With increasing asymmetry the direction of the unstable modes (arrows) 
in the ($\delta\rho_n,\delta\rho_p$) plane
is more and more diverging from the constant concentration value
(thick lines), towards a less asymmetric liquid phase. The
angle between the two directions, i.e. the amount of isospin distillation,
will be proportional to the repulsion of the symmetry term at sub-saturation
densities. 

\subsection*{Symmetry effects at high baryon density: collective flows}
It is
quite desirable to get information on the symmetry energy at higher
density, where furthermore we cannot have complementary investigations
from nuclear structure like in the case of the
low density behaviour. Heavy Ion Collisions ($HIC$) provide a unique
way to create asymmetric matter at high density in terrestrial
laboratories.

The isospin dependence of collective flows has been already discussed
in a non-relativistic framework \cite{flows,bao00}. 
The main new result shown here, in a  Relativistic Mean Field ($RMF$)
scheme \cite{se86}, is the importance at higher energies 
of the microscopic covariant structure 
of the effective interaction in the isovector channel: effective forces 
with very similar symmetry 
terms can give rise to very different flows in relativistic heavy ion
collisions \cite{grefl}.

A full description of the isovector channel in a relativistic 
framework in principle
should
rely on the balance between a scalar ($\delta-like$, attractive) and 
a vector (repulsive) \cite{kubis,liu02,gre03,hole01} contributions. 
This is a quite 
controversial point.
In relativistic $HIC's$, due to the large counterstreaming
nuclear currents, one may directly exploit the
different Lorentz nature of a scalar and a vector field \cite{grefl}.

For the description of heavy ion collisions we solve
the covariant transport equation of the Boltzmann type within the 
Relativistic Landau
Vlasov ($RLV$) method \cite{fuchs95} (for the Vlasov part) and applying
a Monte-Carlo procedure for the collision term, including inelastic 
processes involving
the production/absorption of nucleon resonances, 
\cite{hub94}.
\begin{figure}
\begin{minipage}{75mm}
\begin{center}
\includegraphics*[scale=0.35]{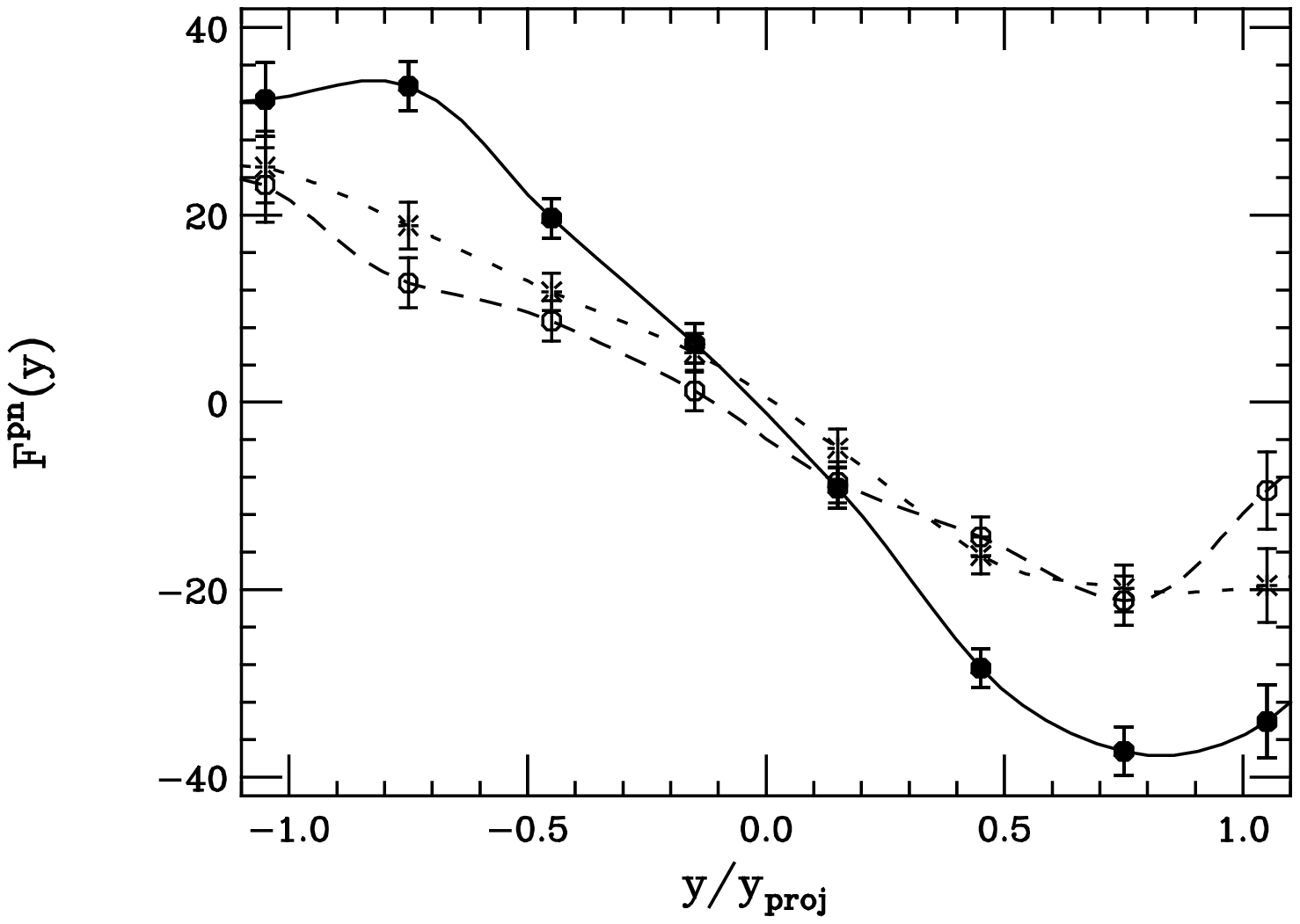}
\caption{Proton-neutron differential collective flow in the
$^{132}Sn+^{132}Sn$
reaction at 1.5 AGeV b=6fm for the three different model for the
isovector mean fields.
Full circles and solid line: $RMF-(\rho+\delta)$.
Open circles and dashed line: $RMF-\rho$.
Stars and short dashed line : $RMF-D\rho$.
}
\label{flows2}
\end{center}
\end{minipage}
\hspace{\fill}
\begin{minipage}{75mm}
\begin{center}
\includegraphics*[scale=0.35]{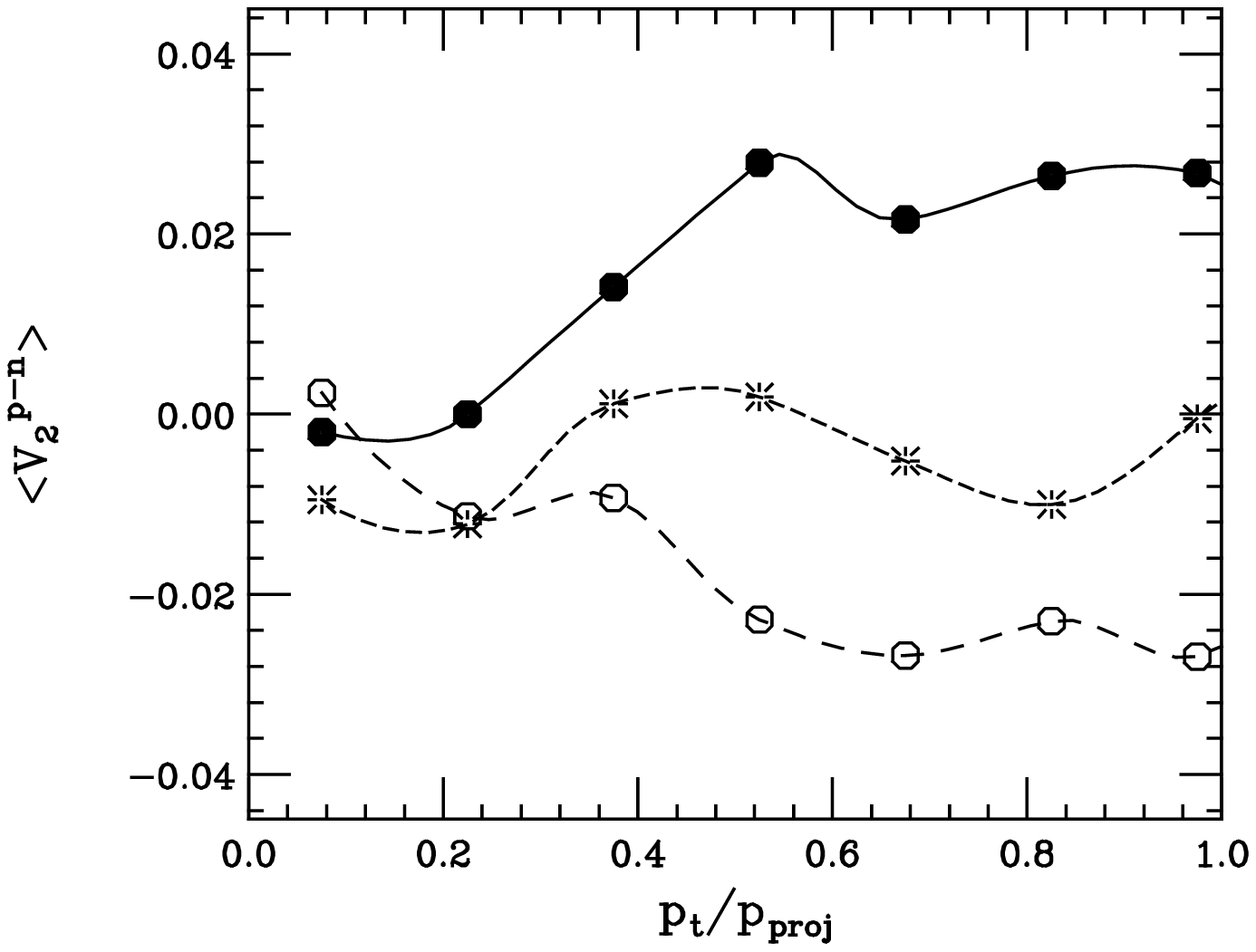}
\caption{Difference between neutron and proton elliptic flow
as a function of the transverse momentum in the
$^{132}Sn+^{132}Sn$ reaction at 1.5 AGeV b=6fm in the
rapidity range $-0.3 \leq y/y_{proj} \leq 0.3$.
Full circles and solid line: $RMF-(\rho+\delta)$.
Open circles and dashed line: $RMF-\rho$.
Stars and short dashed line: $RMF-D\rho$.
}
\label{flows3}
\end{center}
\end{minipage}
\end{figure}

Typical results for the $^{132}Sn+^{132}Sn$ reaction at $1.5AGeV$ 
 (semicentral collisions) are
shown in Figs. \ref{flows2}, \ref{flows3}. In Fig.\ref{flows2}
we report the differential flow
$
F^{pn}(y)\equiv {1/N(y)} \sum_{i} p_{x_{i}} \tau_i$
where $N(y)$ is the total number of free nucleons at the
rapidity $y$, $p_{x_{i}}$ is the transverse momentum of
particle $i$ in the reaction plane, and $\tau_i$ is +1 and -1
for protons and neutrons, respectively.
The
$RMF-(\rho+\delta)$ case (full circles and solid line) presents a stiffer
behaviour relative to the $RMF-\rho$ (open circles) model,
as expected from the more repulsive 
symmetry energy $E_{sym}(\rho_B)$ at high baryon densities \cite{liu02,grefl}. 
We have however repeated the calculation using the
$RMF-D\rho$ interaction, i.e. with only a $\rho$ contribution {\it but}
tuned to reproduce the same $EOS$ of the $RMF-(\rho+\delta)$ case.
The results, short-dashed curve of Fig.\ref{flows2}, are very similar to the
ones of the $RMF-\rho$ interaction. 
Therefore we can explain the large flow effect as mainly due to 
the different strengths of the vector-isovector field 
between $RMF-(\rho+\delta)$
 and $RMF-\rho,D\rho$ in the relativistic dynamics. 
In fact if a source is moving the
vector field is enhanced (essentially by the local $\gamma$
Lorentz factor)
relative to the scalar one. 
Keeping in mind that
$RMF-(\rho+\delta)$
has a three times larger $\rho$ field it is clear that dynamically
the vector-isovector
mean field acting during the $HIC$ is much greater than the one of the
$RMF-\rho,D\rho$ cases.

In Fig.\ref{flows3} we report the elliptic flow $v_{2}(y,p_t)$, 
 $v_2=<(p^2_x-p^2_y)/p^2_t>$
where $p_t=\sqrt{p^2_x+p^2_y}$ is the transverse momentum \cite{daniel}.
A negative value of $v_2$ corresponds to the emission of
matter
perpendicular to the reaction plane, $sqeeze-out$ flow. 
The $p_t$-dependence of
$v_2$ is
very sensitive to the high density behavior of the $EOS$ since highly
energetic
particles ($p_t \ge 0.5$) originate from the initial compressed and
out-of-equilibrium phase of the collision.
We focus on
the proton-neutron difference of the elliptic flow. 
From Fig.\ref{flows3} we see that in the $(\rho+\delta)$
dynamics the high-$p_t$ neutrons show a much larger $squeeze-out$.
This is fully consistent with an early emission (more spectator shadowing)
due to the larger repulsive $\rho$-field. 
The $v_2$ observable, which is a good {\it chronometer} of the reaction
dynamics, appears to be particularly sensitive to the Lorentz structure
of the effective interaction.

\noindent
{\it $\pi^-/\pi^+$ ratios}

\begin{figure}
\begin{center}
\includegraphics*[scale=0.50]{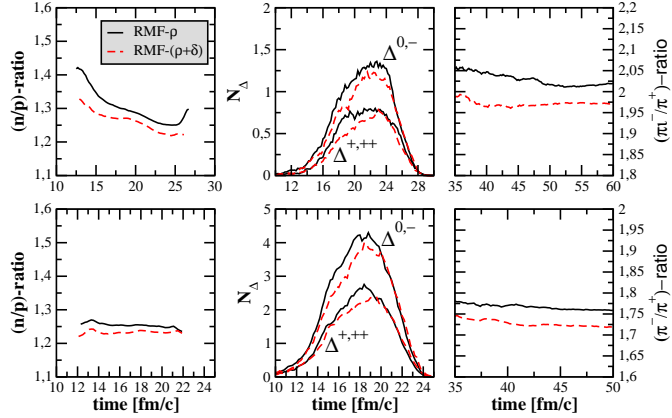}
\caption{Central $Au+Au$ collisions at $0.6AGeV$ (upper) and $1.0AGeV$
(bottom). Time evolution of $n/p$ ratio and $\Delta$ resonance production
in high density regions ($\rho/\rho_0 \ge 2.0$) (first two columns) and
of the total $\pi^-/\pi^+$ ratio (right).
Solid lines: $RMF-\rho$. Dashed lines: $RMF-(\rho+\delta)$.
}
\label{pion}
\end{center}
\end{figure}

Using the same relativistic transport code we have evaluated the
$\pi^-$ vs. $\pi^+$ production for central $Au+Au$ collisions at different 
energies, see Fig.\ref{pion}.
As expected the larger repulsion seen by neutrons at high densities
in the $RMF-(\rho+\delta)$ will show up in a reduced $n/p$ ratio,
smaller $\Delta^{0,-}$ density and finally a reduced $\pi^-$
production. However now reabsorption effects are important and
actually the effect appears to be decreasing at higher energies.

\subsection*{Isospin and Deconfinement at High Baryon Density}
It is relatively easy to show that at high baryon density and low
temperature we can expect a transition from hadronic matter to deconfined
quark matter.
The procedure is straightforward:

\noindent
(i) Start from two "reasonable" model Equations of State ($EOS$), one
for the hadronic phase, which can reproduce saturation properties, one
for the quark phase, which can reproduce the hadron spectrum.

\noindent
(ii)Construct the phase separation boundary surface from the Gibbs
phase rule.

For symmetric matter the baryon density $\rho_{tr}$ corresponding
to the transition to the coexistence region is relatively high, as 
expected, ranging from 4 to 8 times the saturation value $\rho_0$,
depending on the stiffness of the hadronic $EOS$ at high densities.
The new feature we would like to focus on in this report is the isospin
dependence of such boundary location. We can foresee an interesting 
asymmetry effect, in the appealing direction of a decrease of $\rho_{tr}$,
since the hadronic $EOS$ becomes more repulsive ref.\cite{mue97,isoquark}  .

\begin{figure}
\begin{center}
\includegraphics*[scale=0.37]{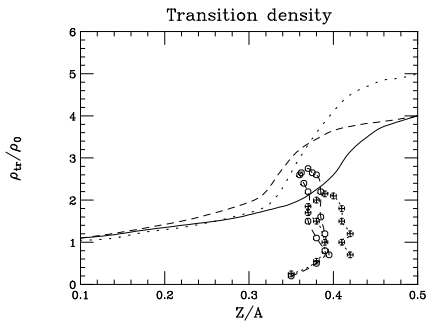}
\caption{\it 
Variation of the transition density with proton fraction
followed in the interaction zone during a semicentral $^{132}Sn 
+ ^{132}Sn$ collision at 1AGeV (circles) and 300AMeV (crosses)} 
\label{rhocr}
\end{center}
\end{figure}

The proton fraction $Z/A$ dependence of the $\rho_{tr}$
is reported in Fig.\ref{rhocr} with the bag constant 
value $B^{1/4}=150$ MeV 
and $\alpha_s=0$ for
the quark phase and various choices for the
hadronic $EOS$: {\it Dotted~line} $GM3$ parametrization \cite{gle92};
{\it Dashed~line} $RMF-\rho$ parametrization \cite{liu02};
{\it Solid~line} $RMF-(\rho+\delta)$ parametrization \cite{liu02}.
$GM3$ and $RMF-\rho$ have the same source of the interaction symmetry term
(only the $\rho$-meson). 
The effects of the asymmetry appears now quite dramatic: we
see a $\rho_{trans}$ as low as $2\rho_0$ for proton fractions between
0.3 and 0.4, conditions that with some confidence we could "locally"
reach in a heavy ion collision at {\it intermediate} energy using
exotic  very asymmetric beams.

Using our Relativistic Transport Code, with the $RMF-(\rho+\delta)$ 
effective interaction, we have performed some simulations
of the $^{132}Sn~+~^{132}Sn$ (average $Z/A=0.38$) collision at various 
energies, for a semicentral impact parameter, $b~=~6fm$, just to optimize the
neutron skin effect. In Fig.\ref{rhocr} the paths in the $(\rho,Z/A)$ plane 
followed in the c.m. region
during the collision are reported, at two energies $300~AMeV$ (crosses)
and $1~AGeV$ (circles). We see that already at $300~AMeV$ we are reaching
the border of the mixed phase, and we are well inside it at $1~AGeV$.

In conclusion we support the possibility of observing precursor signals
of the phase transition to a deconfined matter in violent collision
(central and semicentral) of exotic (radioactive) heavy ions
in the energy range of few hundred MeV per nucleon.
A possible signature could be revealed through an earlier "softening"
of the hadronic $EOS$ for larger asymmetries.

\subsection*{Outlook: The Eleven Observables}
As a conclusion of our report we like to suggest a selection
of {\it Eleven Observables}, from low to relativistic energies, that 
we expect particularly sensitive
to the microscopic structure of the {\it in medium }interaction
in the isovector channel, i.e. to the symmetry energy and its
``fine structure'':

\noindent
{\it 1. Competition of Reaction Mechanisms}. Interplay of low-energy
 dissipative mechanisms, e.g. \cite{fab98}, fusion (incomplete) vs.
 deep-inelastic vs. neck fragmentation: a stiff symmetry term leads to
 a  more repulsive dynamics.

\noindent
{\it 2. N/Z of fast nucleon emission}. Symmetry repulsion of the
 neutron/proton mean field in various density regions.

\noindent
{\it 3. Neutron/Proton correlation functions}. Time-space structure
 of the fast particle emission and its relation to the baryon density
 of the source, see the recent \cite{chen02}.

\noindent
{\it 4. Fragment Multiplicities}. A more efficient use of protons 
in forming primary fragments is expected in the asy-stiff case.

\noindent
{\it 5. Isospin Distillation (Fractionation)}. Isospin content
of the Intermediate Mass Fragments in central collisions. Test of 
the symmetry term in dilute matter.

\noindent
{\it 6. Isospin content of Neck-Fragments}. Test of the 
symmetry term around $\rho_0$.

\noindent
{\it 7. Fast Fission Multiplicity}. The rate of ``aligned''
fission events of the Projectile-Like/Target-Like Fragments reflects
the symmetry repulsion in semicentral collisions.

\noindent
{\it 8. Isospin Diffusion}. Measure of charge equilibration
in the ``spectator'' region in semicentral collisions, test of
symmetry repulsion.

\noindent
{\it 9. Neutron-Proton Collective Flows}. Together with
 light isobar flows. Check of symmetry transport effects.
Test of the momentum dependence (relativistic structure) of the
interaction in the isovector channel.
 Measurements also for different $p_t$ selections. 

\noindent
{\it 10. $\pi^-/\pi^+$ Yields}. Since $\pi^-$ are mostly produced
in $nn$ collisions we can expect a reduction for highly repulsive
symmetry terms at high baryon density, see \cite{fae97,baopi}.

\noindent
{\it 11. Deconfinement Precursors}. Signals of a mixed phase formation
 (quark-bubbles) in high baryon density regions reached with asymmetric 
 $HIC$ at intermediate energies.

From the points $3, 4, 5, 8, 9$ in our simulations we presently get
some indications for {\it asy-stiff} behaviors, i.e. increasing
repulsive density dependence of the symmetry term, but not more
fundamental details. Moreover all the available data are obtained
with stable beams, i.e. within low asymmetries.
\vskip 0.3cm

\noindent
{\it Acknowledgements}:
In this report we are collecting also ideas and results partially
reached in a very pleasant and fruitful collaboration with B.Liu, 
F.Matera, M.Zielinzka-Pfabe', J.Rizzo and H.H.Wolter. We warmly thank all 
of them.

\end{document}